\documentclass[3p,times,procedia,sort&compress]{elsarticle}
\usepackage{nupha_ecrc}

\volume{00}

\firstpage{1}

\journalname{Nuclear Physics A}

\runauth{B. Krouppa, R. Ryblewski, and M. Strickland}

\jid{nupha}

\jnltitlelogo{Nuclear Physics A}

\usepackage{amssymb}
\usepackage{color}
\definecolor{darkblue}{RGB}{0,0,196}
\definecolor{darkred}{RGB}{196,0,0}
\usepackage[colorlinks=true,linktocpage=true,linkcolor=darkblue,citecolor=darkblue,urlcolor=darkblue]{hyperref}

\usepackage[figuresright]{rotating}

\begin{document}

\begin{frontmatter}



%% Instructions from Editor: Please use the following \dochead only in the preprint version (e-print arXiv etc.); 
%% use empty \dochead{} when submitting to Nuclear Physics A!
\dochead{XXVIth International Conference on Ultrarelativistic Nucleus-Nucleus Collisions\\ (Quark Matter 2017)}
%% e.g. \dochead{17th International Conference on Dynamical Processes in Excited States of Solids}

\title{Bottomonium suppression in heavy-ion collisions}


\author[kent]{Brandon Krouppa}
\author[krakow]{Radoslaw Ryblewski}
\author[kent]{Michael Strickland}

\address[kent]{Department of Physics, Kent State University, Kent, OH 44242 United States}
\address[krakow]{The H. Niewodnicza\'nski Institute of Nuclear Physics, Polish Academy of Sciences, PL-31342 Krak\'ow, Poland}

\begin{abstract}
The thermal suppression of heavy quark bound states represents an ideal observable for determining if one has produced a quark-gluon plasma in ultrarelativistic heavy-ion collisions. In recent years, a paradigm shift has taken place in the theory of quarkonium suppression due to new first principles calculations of the thermal widths of these states. These thermal widths are large, e.g. O(20-100 MeV) for the $\Upsilon(1S)$, and cause in-medium suppression of the states at temperatures below their traditionally defined disassociation temperatures. In order to apply the newly developed understanding to phenomenology, however, one must make detailed 3+1d dissipative hydrodynamical models of the plasma including the effects of finite shear viscosity. These effects include not only the modification of the time evolution of the temperature of the system, flow, etc., but also non-equilibrium modifications of the heavy quark potential itself.  In this proceedings contribution, we briefly review the setup for these model calculations and present comparisons of theory with data from RHIC 200 GeV/nucleon Au-Au collisions, LHC 2.76 TeV/nucleon Pb-Pb, and LHC 5.02 TeV/nucleon Pb-Pb collisions as a function of number of participants, rapidity, and transverse momentum.
\end{abstract}

\begin{keyword}


\end{keyword}

\end{frontmatter}

%\linenumbers

\section{Introduction}
\label{sec:introduction}

Ultrarelativistic heavy-ion collision (URHIC) experiments being performed at the Large Hadron Collider (LHC) at CERN and the Relativistic Heavy Ion Collider (RHIC) at Brookhaven National Laboratory aim to create a primordial state of matter called the quark-gluon plasma (QGP).   Comparisons between theory and experiment suggest that LHC URHICs produce a QGP with an initial temperature on the order of $T_{0}=600{-}700$ MeV at $\tau_0 = 0.25$ fm/c \cite{Alqahtani:2017jwl} and analysis of collective flow of the matter produced indicates that the QGP behaves like a nearly inviscid relativistic fluid \cite{Heinz:2013AnnRevNuc,Gale:2013IntJModPhys,Alqahtani:2017jwl}.  In order to further study the properties of the QGP, hard probes such as the production of bottomonium states can be used.  Bottomonia are of particular interest because such states can survive well into the deconfined phase due to their large vacuum binding energies ($\lesssim$ 1 GeV) and the relative production of the ground state and various excited states can be used to infer properties of the QGP such as its initial temperature, shear viscosity to entropy density ratio, etc.  

One theoretical complication faced in modeling bottomonium production and decay in the QGP is that, since bottomonium states are formed early ($\tau < 1$ fm/c), they are particularly sensitive to the early-time non-equilibrium dynamics of the QGP.  Of particular importance is the large pressure anisotropy of the QGP in the local rest frame, ${\cal P}_L \ll {\cal P}_T$, which is induced by the rapid longitudinal expansion of the QGP created in URHICs.  This pressure anisotropy leads to important non-equilibrium corrections to the widths of the various bottomonium states \cite{Dumitru:2007hy,Burnier:2009yu,Dumitru:2009fy,Strickland:2011mw,Strickland:2011aa,Du:2016wdx,Biondini:2017qjh} which we take into account using 3+1d dissipative anisotropic hydrodynamics~\cite{Strickland:2014pga}.  In this proceedings, we present a collection of our model predictions presented originally in Refs.~\cite{Krouppa:2015yoa,Krouppa:2016jcl} along with some of the experimental data available from the CMS, ALICE, and STAR collaborations.  For details of the theoretical framework and methods used, we refer the reader to Refs.~\cite{Strickland:2011aa,Krouppa:2015yoa,Krouppa:2016jcl}.

%%%%%%%%%%%%%%%%%%%%%%%%%%%%%%%%%%%%%%%%%%%%%%%%%%%%%%%%%%%
\begin{figure}[t!]
\centerline{
\includegraphics[width=0.485\linewidth]{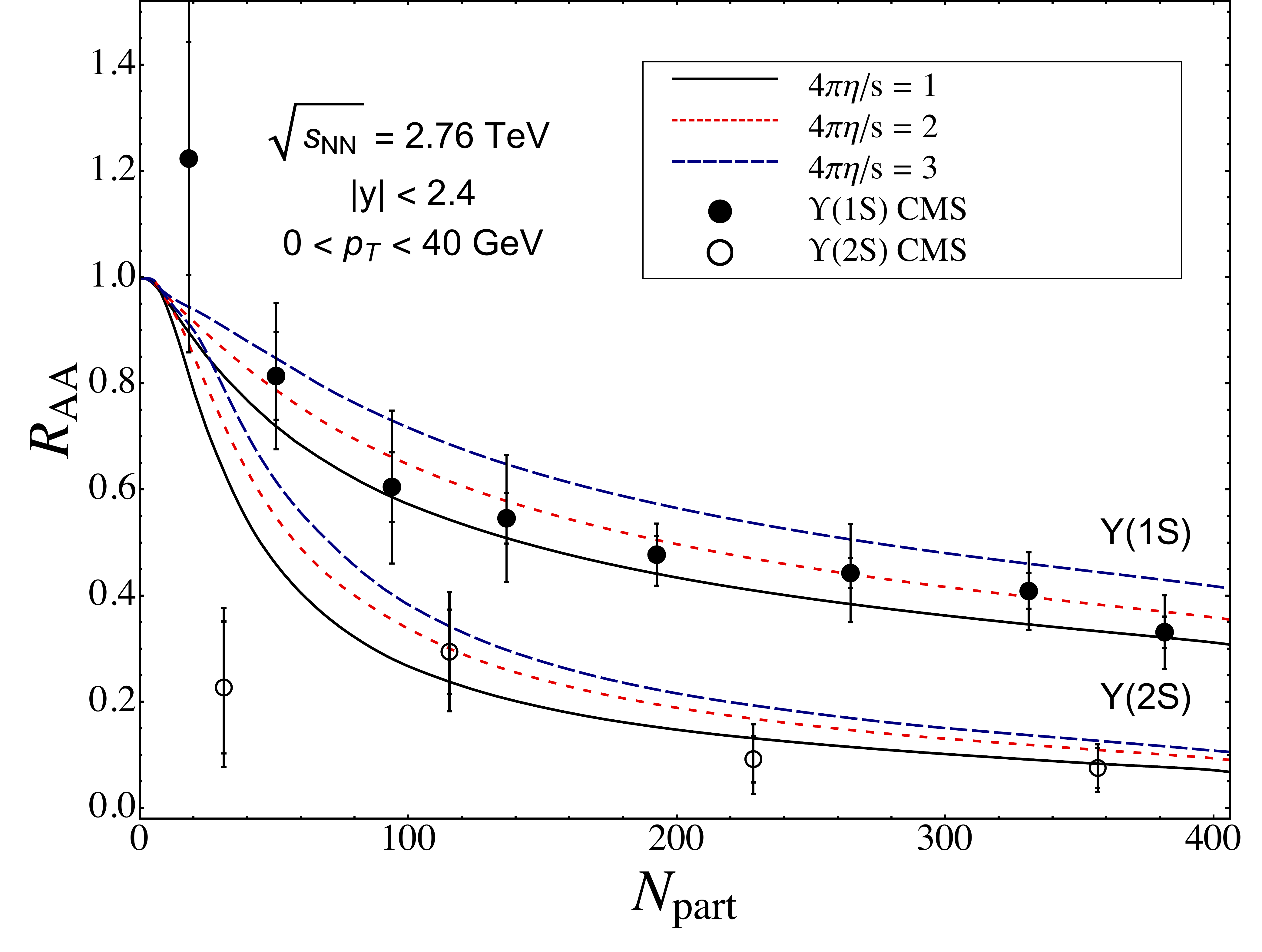}
\includegraphics[width=0.485\linewidth]{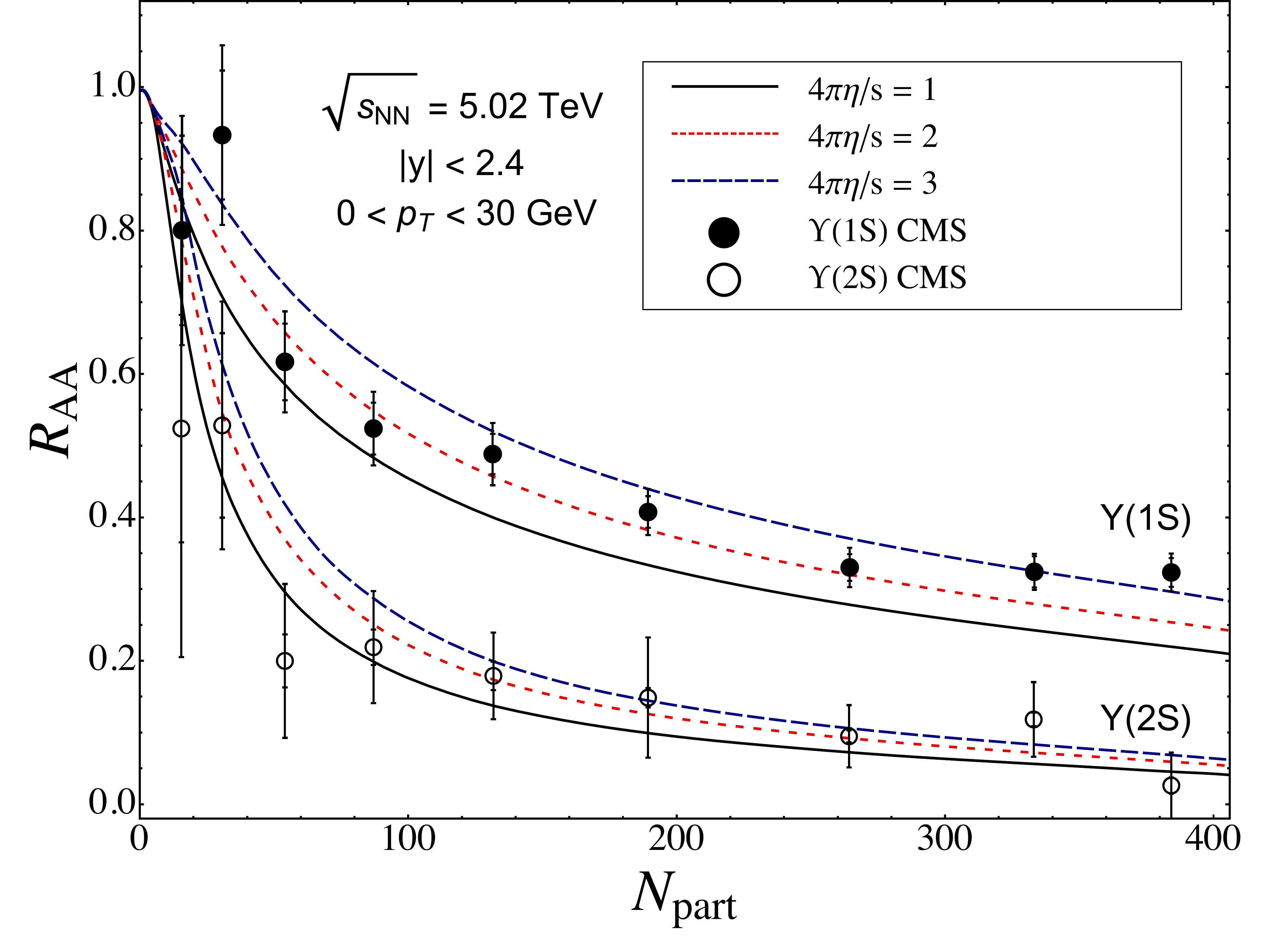}
}
\caption{$R_{AA}$ versus $N_{\rm part}$ for $\Upsilon(1S)$ and $\Upsilon(2S)$.  Left panel shows $\sqrt{s_{NN}} = 2.76$ TeV with CMS data from \cite{Khachatryan:2016xxp} and the right panel shows $\sqrt{s_{NN}} = 5.02$ TeV with preliminary CMS data from \cite{CMS5TeV}.  Lines correspond to three different values of the shear viscosity to entropy density ratio.}
\label{fig:1}
\end{figure}
%%%%%%%%%%%%%%%%%%%%%%%%%%%%%%%%%%%%%%%%%%%%%%%%%%%%%%%%%%%

%%%%%%%%%%%%%%%%%%%%%%%%%%%%%%%%%%%%%%%%%%%%%%%%%%%%%%%%%%%
\begin{figure}[t!]
\centerline{
\includegraphics[width=0.485\linewidth]{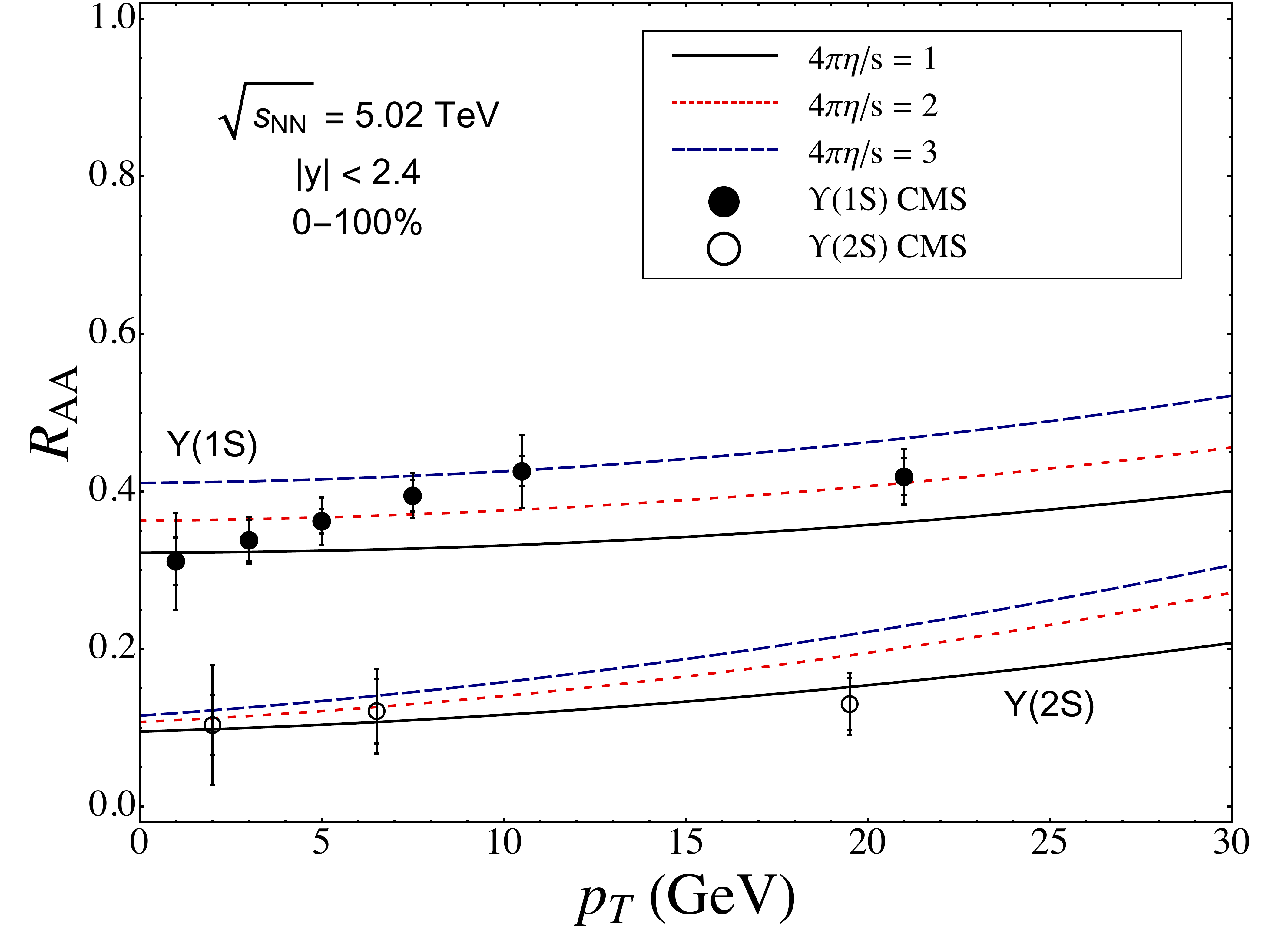}
\includegraphics[width=0.485\linewidth]{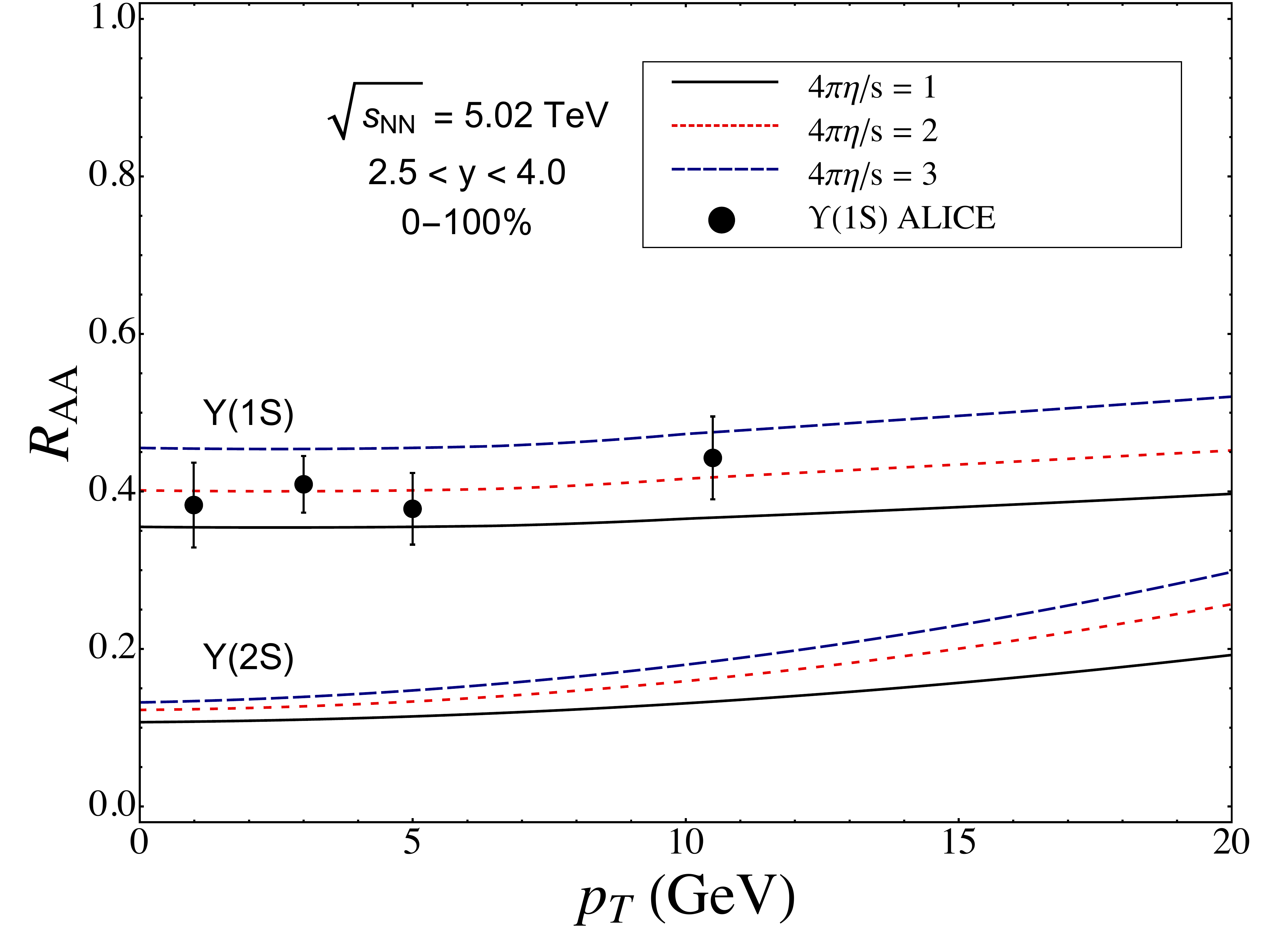}
}
\caption{$R_{AA}$ versus $p_T$ for $\Upsilon(1S)$ and $\Upsilon(2S)$ at $\sqrt{s_{NN}} = 5.02$ TeV.  Left panel shows preliminary CMS data from \cite{CMS5TeV} and the right panel shows preliminary ALICE data from \cite{ALICE5TeV}.  Lines are same as Fig.~\ref{fig:1}.}
\label{fig:2}
\end{figure}
%%%%%%%%%%%%%%%%%%%%%%%%%%%%%%%%%%%%%%%%%%%%%%%%%%%%%%%%%%%

%%%%%%%%%%%%%%%%%%%%%%%%%%%%%%%%%%%%%%%%%%%%%%%%%%%%%%%%%%%
\begin{figure}[t!]
\centerline{
\includegraphics[width=0.485\linewidth]{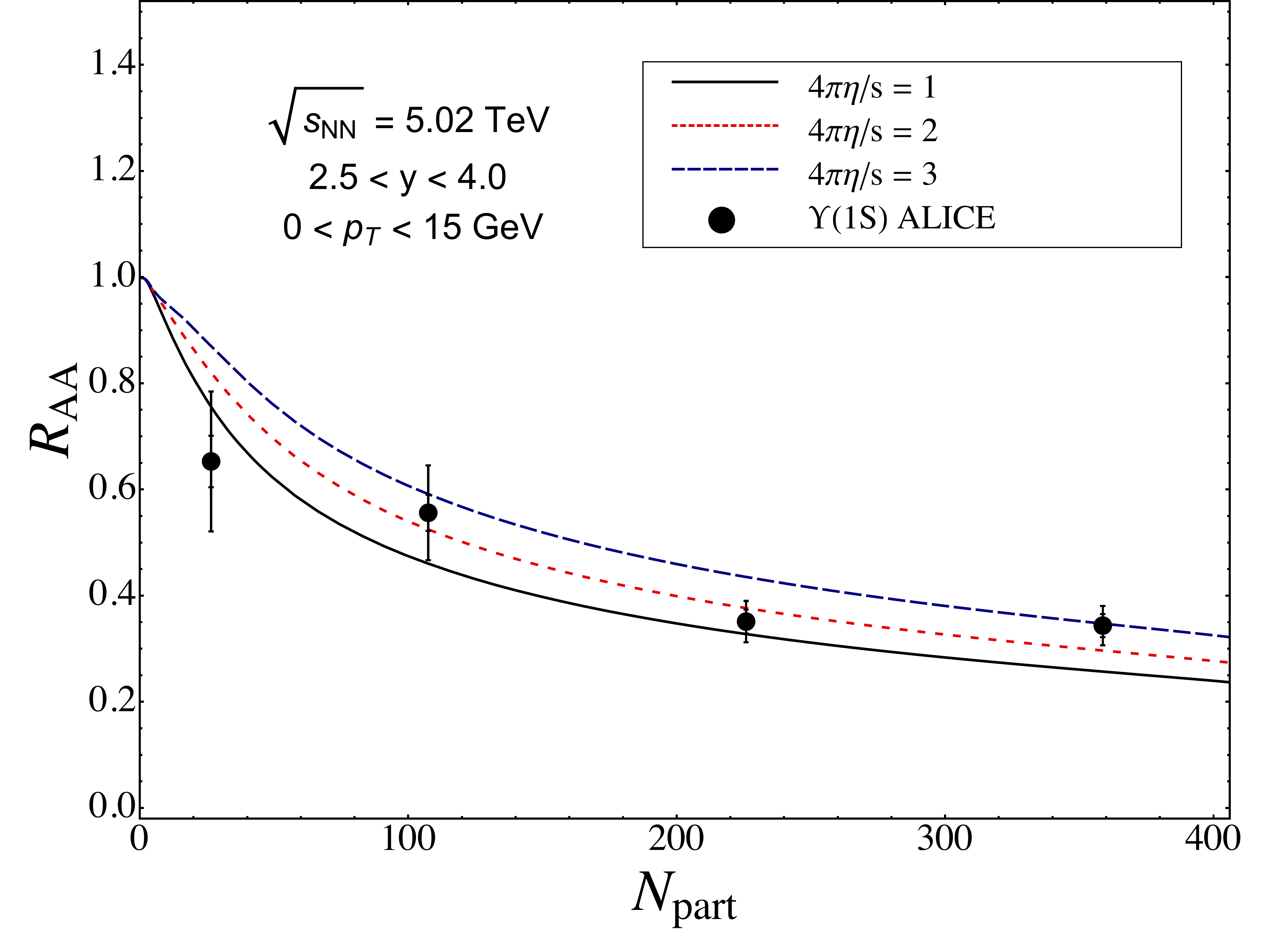}
\includegraphics[width=0.485\linewidth]{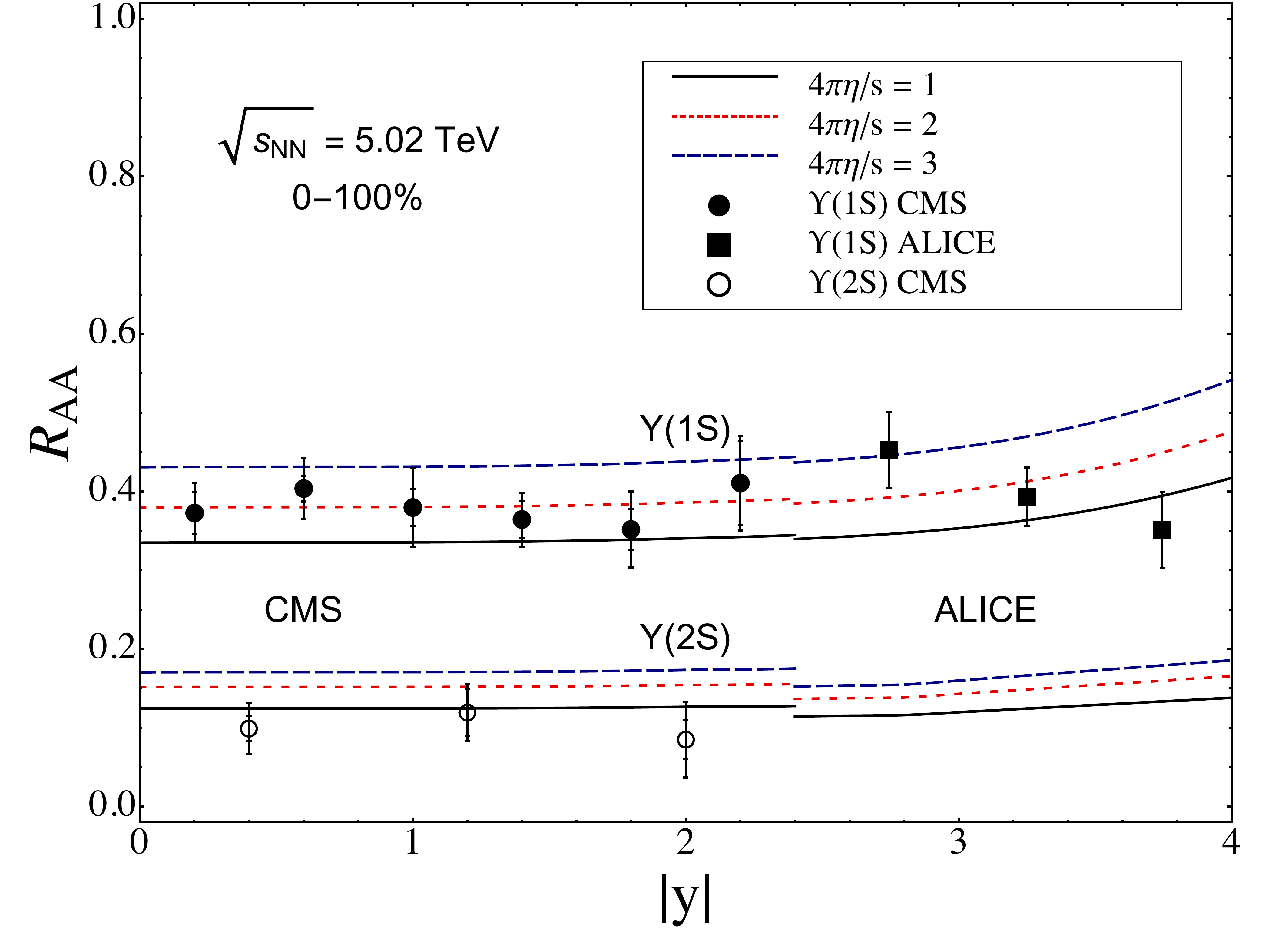}
}
\caption{(Left) $R_{AA}$ versus $N_{\rm part}$ for $\Upsilon(1S)$ at $\sqrt{s_{NN}} = 5.02$ TeV compared to preliminary ALICE data \cite{ALICE5TeV}. (Right) $R_{AA}$ versus $y$ for $\Upsilon(1S)$ at $\sqrt{s_{NN}} = 5.02$ TeV.  The right panel figure is split at $y=2.4$, with the CMS $p_T$-cut applied for $y<2.4$ and the ALICE $p_T$-cut applied for $y>2.4$.
}
\label{fig:3}
\end{figure}
%%%%%%%%%%%%%%%%%%%%%%%%%%%%%%%%%%%%%%%%%%%%%%%%%%%%%%%%%%%

%%%%%%%%%%%%%%%%%%%%%%%%%%%%%%%%%%%%%%%%%%%%%%%%%%%%%%%%%%%
\begin{figure}[t!]
\centerline{
\includegraphics[width=0.485\linewidth]{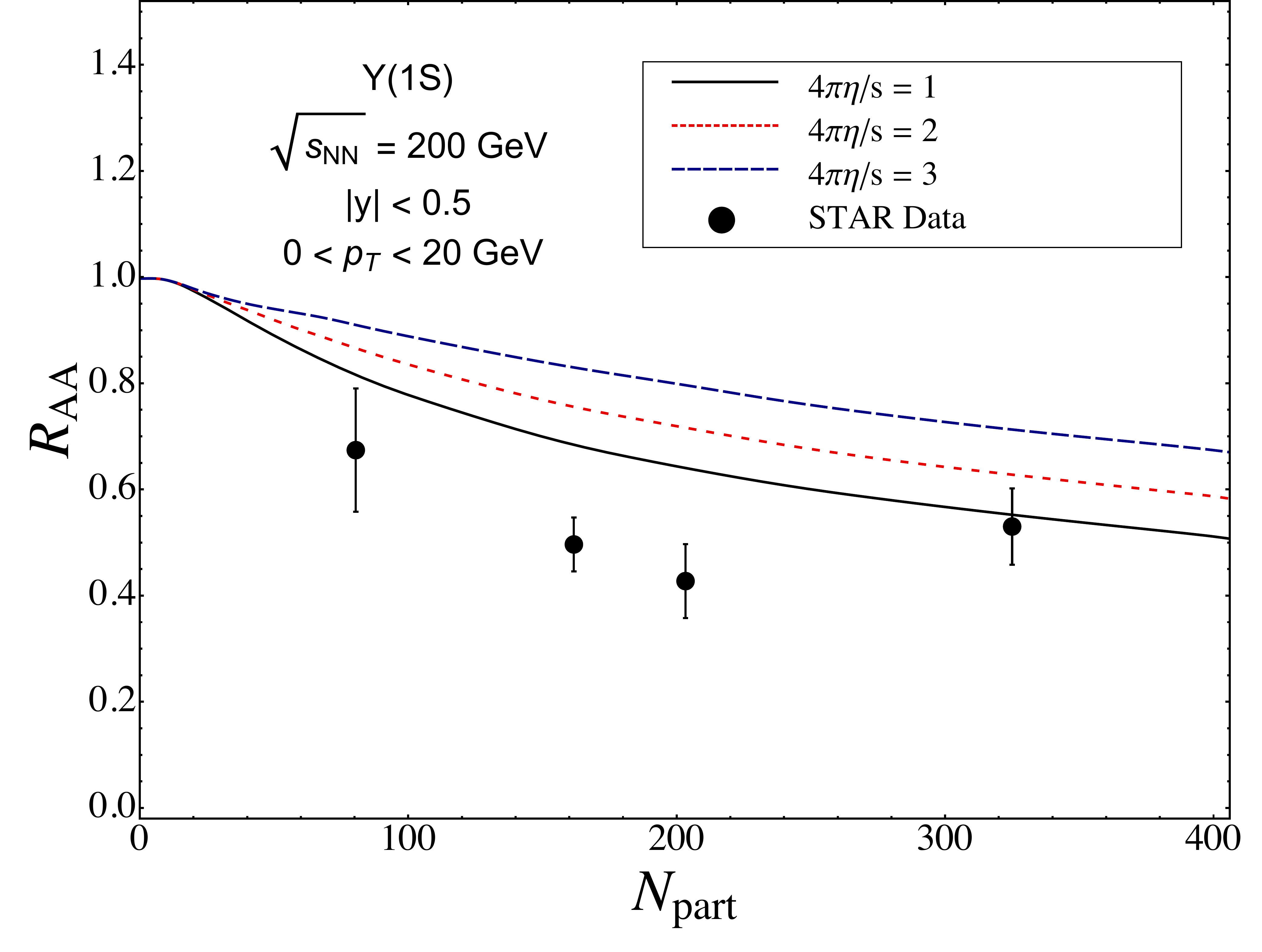}\hspace{2mm}
\includegraphics[width=0.485\linewidth]{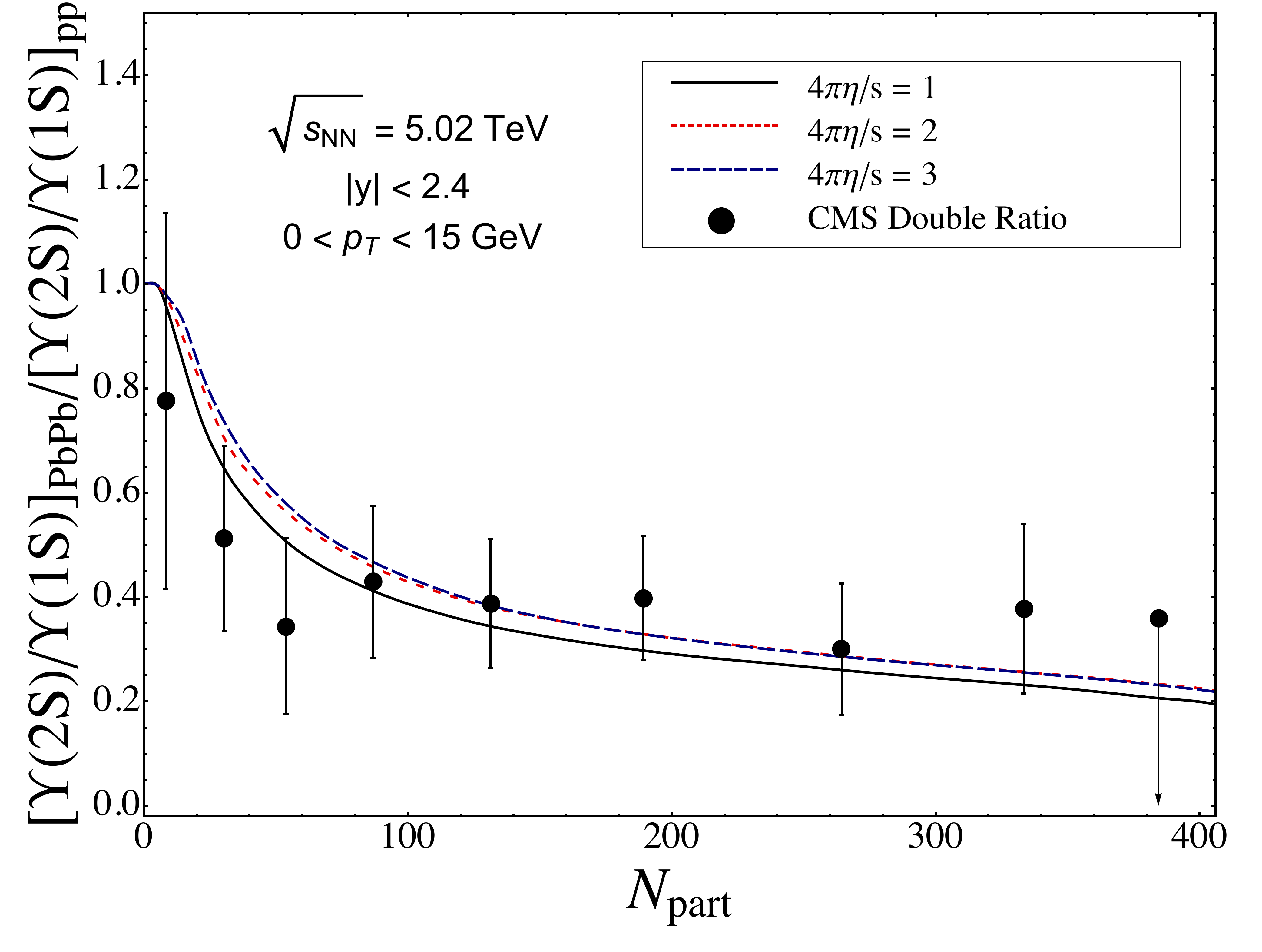}
}
\caption{(Left) $R_{AA}$ versus $N_{\rm part}$ for $\Upsilon(1S)$ at $\sqrt{s_{NN}} = 200$ GeV compared to preliminary STAR data \cite{STARupsilon}. (Right) $[\Upsilon(2S)/\Upsilon(1S)]_{\rm PbPb}/[\Upsilon(2S)/\Upsilon(1S)]_{\rm pp}$ vs $N_{\rm part}$ at $\sqrt{s_{NN}} = 5.02$ TeV compared to preliminary CMS data.
}
\label{fig:4}
\end{figure}
%%%%%%%%%%%%%%%%%%%%%%%%%%%%%%%%%%%%%%%%%%%%%%%%%%%%%%%%%%%

\section{Results}

We now turn to our results.  In Fig.~\ref{fig:1}, we present $R_{AA}$ versus $N_{\rm part}$ for $\Upsilon(1S)$ and $\Upsilon(2S)$.  The left panel shows $\sqrt{s_{NN}} = 2.76$ TeV model predictions together with CMS data from \cite{Khachatryan:2016xxp} and the right panel shows $\sqrt{s_{NN}} = 5.02$ TeV with preliminary CMS data from \cite{CMS5TeV}.  The lines correspond to three different values of the shear viscosity to entropy density ratio.  For all three theoretical curves shown in all plots we assumed that the system was initially isotropic in momentum space and, for each value of $4\pi\eta/s$, we tuned the initial central temperature such that the final soft hadron multiplicity was held fixed.  For $\sqrt{s_{NN}} = 2.76$ TeV, this resulted in $T_0 = \{ 0.552, 0.546, 0.544 \}$ GeV and, for $\sqrt{s_{NN}} = 5.02$ TeV, this resulted in $T_0 = \{ 0.641, 0.632, 0.629 \}$ GeV when taking $4\pi\eta/s  = \{ 1,2,3 \}$ .  In both cases, we initialized the hydrodynamic evolution at $\tau_0 = 0.3$ fm/c.  As we can see from Fig.~\ref{fig:1}, the model does a very good job compared to CMS preliminary data at both collision energies.

In Fig.~\ref{fig:2}, we present $R_{AA}$ versus $p_T$ for $\Upsilon(1S)$ and $\Upsilon(2S)$ at $\sqrt{s_{NN}} = 5.02$ TeV.  The left panel shows preliminary CMS data from \cite{CMS5TeV} and the right panel shows preliminary ALICE data from \cite{ALICE5TeV}.  In Fig.~\ref{fig:3}, we present $R_{AA}$ versus $N_{\rm part}$ for $\Upsilon(1S)$ at $\sqrt{s_{NN}} = 5.02$ TeV compared to preliminary ALICE data in the left panel and $R_{AA}$ versus $y$ for $\Upsilon(1S)$ compared to both CMS and ALICE results.  In all cases, we find quite reasonable agreement with both the CMS and ALICE $\sqrt{s_{NN}} = 5.02$ TeV results.   Finally, in the left panel of Fig.4 we present the result obtained for $\Upsilon(1S)$ suppression in $\sqrt{s_{NN}} = 200$ GeV Au-Au collisions compared to preliminary STAR data.  As we can see from this panel, the model does not seem to predict the same amount of suppression as seen by STAR.  This could be due to the relative importance of cold nuclear matter effects at RHIC energies or could point to energy dependence of the feeddown fractions in the Upsilon sector.  This clearly deserves further attention.  In the right panel of Fig.~\ref{fig:4}, we present our results for the double ratio $[\Upsilon(2S)/\Upsilon(1S)]_{\rm PbPb}/[\Upsilon(2S)/\Upsilon(1S)]_{\rm pp}$ compared to preliminary experimental data from CMS.  This observable is particularly nice, because many systematic errors cancel in this ratio.  As can be seen from this final figure, at $\sqrt{s_{NN}} = 5.02$ TeV our model provides a quite reasonable description of the preliminary CMS results. 

\section{Discussion and conclusions}

In this proceedings contribution we have provided a collection of our results related to bottomonium suppression at both RHIC and LHC energies along with a comparison to experimental data.  The model utilized includes the effects of in-medium decays through the use of a complex-valued potential which has been extended to non-equilibrium \cite{Dumitru:2007hy,Burnier:2009yu,Dumitru:2009fy,Strickland:2011mw,Strickland:2011aa}.  We did not include cold nuclear matter effects or regeneration in the model.  The in-medium decay widths obtained were folded over the full 3+1d anisotropic hydrodynamics (aHydro) evolution of the QGP in order to compute the survival probability of the various Upsilon states as a function of space-time position, transverse momentum, and rapidity.  

We find that the model does a quite reasonable job in describing the CMS $R_{\rm AA}$ data for $\Upsilon(1S)$ and $\Upsilon(2S)$ at both $\sqrt{s_{NN}} = 2.76$ TeV and 5.02 TeV as a function $N_{\rm part}$, $p_T$, and $y$.  In addition, we find that the model is compatible with CMS's observed 2S to 1S ratio at $\sqrt{s_{NN}} = 5.02$ TeV.  At forward rapidity, we also find that the model is compatible with the ALICE $R_{\rm AA}$ data for $\Upsilon(1S)$ at $\sqrt{s_{NN}} = 5.02$ TeV as a function $N_{\rm part}$ and $p_T$.  That said, the ALICE data suggest that $R_{AA}$ might be an decreasing function of rapidity at large rapidity, however, when combined with the CMS results at mid-rapidity, the variation seen is consistent with a flat rapidity dependence (see Fig.~\ref{fig:3} right panel).  Finally, there is some tension with the STAR data and further work is needed to assess the role of cold nuclear matter effects at these energies and the possibility of energy dependence of the mixing fractions.  In addition, in the future we plan to upgrade the background evolution to include non-conformal effects such as bulk viscous corrections to the evolution \cite{Alqahtani:2017jwl} and to also include bulk corrections to the heavy-quark potential itself \cite{Du:2016wdx}.

\vspace{3mm}

\noindent
{\bf Acknowledgments}:  B.~Krouppa and M.~Strickland were supported by U.S. DOE Award No.~DE-SC0013470.  R.~Ryblewski was supported by Polish National Science Center grant No. DEC-2012/07/D/ST2/02125.

\vspace{-2mm}

%%%%%%%%%%%%%%%%%%%%%%%%%%%%%%%%%%%%%%%%%%%%%%%%%%%%%%%%%%%



%% References


\bibliographystyle{utphys}
\bibliography{strickland}






% \end{thebibliography}

\end{document}